\documentclass[english,utf8,twocolumn,a4paper]{aa}
\usepackage[T1]{fontenc}
\usepackage{color}
\usepackage{graphicx}
\usepackage{ulem}
\usepackage{graphicx}
\usepackage{footnote}
\usepackage{csquotes}

\definecolor{pad}{rgb}{0.77,0.07,0.77}
\usepackage{t1enc}
\usepackage{times}
\usepackage{ulem}

\makeatletter

\providecommand{\tabularnewline}{\\}

\def\farcs{\hbox{$.\!\!^{\prime\prime}$}}

\makeatother

\usepackage{babel}
\begin{document}

\title{Have we missed an interstellar comet four years ago?}

\author{Piotr A. Dybczy\'{n}ski\textsuperscript{1} and Ma{\l}gorzata Kr{\'o}likowska\textsuperscript{2}}

\institute{\textsuperscript{1}Astronomical Observatory Institute, Faculty of
Physics, A.Mickiewicz Univ, Pozna\'{n}, Poland, e-mail: dybol@amu.edu.pl,
\textsuperscript{2}Space Research Centre, Polish Academy of Sciences,
Warsaw, Poland, e-mail: mkr@cbk.waw.pl}

\abstract{New orbit for C/2014~W$_{10}$ PANSTARRS is obtained. High original
eccentricity of $e=1.65$ might suggest an interstellar origin of
this comet. The probable reasons for missing this possible important
event is discussed and a call for searching potential additional observations
in various archives is proposed.}
\maketitle

\section{Introduction}

In the course of a redetermination of orbits of long period comets
(hereafter LPCs) we recently attempted to check, whether the assumption
of a parabolic orbit for hundreds of comets discovered after 1950
is fully justified in all cases (Kr{\'o}likowska \& Dybczy\'{n}ski, in
preparation). During this research we found an interesting case of
the comet C/2014~W$_{10}$PANSTARRS. It was observed during one month
only, with a small total number of 17 observations distributed very heterogeneously
(November~25--28 and December~22).  The data is extremely scarce, however it surprised us that the published orbital solutions for this comet are  so different.

The discovery of C/2014~W$_{10}$ was announced by R.~Wainscoat and
R.~Weryk in a Minor Planet Center CBET~4030 electronic telegram on
the 5\textsuperscript{th}of December, 2014. This comet was discovered
on the 25\textsuperscript{th} of November 2014 in four w-band CCD
exposures taken with the 1.8-m Pan-STARRS1 telescope. During the follow-up
observations the recognizable coma of 1-8 arcsec level was reported,
as well as the barely visible tail. Full details might be found in
the quoted CBET\footnote{http://www.cbat.eps.harvard.edu/cbet/004000/CBET004030.txt (subscription
needed)}.

Basing on 14 observations taken during three days: November 25, 26
and 28 the preliminary orbit was published in the quoted CBET and
the following MPC electronic circular: MPEC~2014-X31, issued on the
same day. At MPC they obtained the elliptical, short period orbit
with elements presented in the first data column in Table \ref{tab:various_preliminary_orbits}.

\begin{table*}[h]
\caption{\label{tab:various_preliminary_orbits}Publicly available C/2014~W$_{10}$
orbits from various sources. Angular elements with respect to the
ecliptic of J2000. Only for the orbit given at JPL we were able to reproduce also the uncertainties of orbital elements.}

\begin{tabular}{cccccc}
\hline 
{\scriptsize{}Orbit calculator} & {\scriptsize{}MPC prelim.} & {\scriptsize{}MPC final} & {\scriptsize{}JPL} & {\scriptsize{}Kinoshita} & {\scriptsize{}Patrick Rocher}\tabularnewline
\hline 
{\scriptsize{}number of observations} & {\scriptsize{}14} & {\scriptsize{}17} & {\scriptsize{}17} & {\scriptsize{}14} & {\scriptsize{}17}\tabularnewline
{\scriptsize{}mean residual} & {\scriptsize{}n.a.} & {\scriptsize{}n.a.} & {\scriptsize{}0\farcs24} & {\scriptsize{}0\farcs19} & {\scriptsize{}0\farcs21}\tabularnewline
 
{\scriptsize{}perihelion passage date {[}TT{]}} & {\scriptsize{}2015 Feb. 9.246} & {\scriptsize{}2014 Feb. 06.757} & {\scriptsize{}2014 Feb. 07.6 $\pm$ 979 days} & {\scriptsize{}2016 Aug. 17.267 } & {\scriptsize{}2013 Aug. 29.115}\tabularnewline
 
{\scriptsize{}perihelion distance {[}AU{]}} & {\scriptsize{}7.4247577} & {\scriptsize{}7.9952199} & {\scriptsize{}8.00 $\pm$ 0.75} & {\scriptsize{}6.51927} & {\scriptsize{}7.7604333}\tabularnewline

{\scriptsize{}eccentricity} & {\scriptsize{}0.6039453 } & {\scriptsize{}1.0} & {\scriptsize{}1.0 $\pm$ 0.87} & {\scriptsize{}1.0} & {\scriptsize{}1.2233709}\tabularnewline

{\scriptsize{}semimajor axis {[}AU{]}} & {\scriptsize{}18.7467988} & {\scriptsize{}\textendash{}} & {\scriptsize{}\textendash{}} & {\scriptsize{}\textendash{}} & {\scriptsize{}\textendash{}}\tabularnewline

{\scriptsize{}inclination {[}deg{]}} & {\scriptsize{}72.97258} & {\scriptsize{}85.11344} & {\scriptsize{}85.13 $\pm$ 12} & {\scriptsize{}76.637} & {\scriptsize{}87.760905}\tabularnewline

{\scriptsize{}longitude of the ascending node {[}deg{]}} & {\scriptsize{}40.86266} & {\scriptsize{}43.68179} & {\scriptsize{}43.68 $\pm$ 2.8} & {\scriptsize{}41.695} & {\scriptsize{}44.289828}\tabularnewline

{\scriptsize{}argument of perihelion {[}deg{]}} & {\scriptsize{}19.01185} & {\scriptsize{}356.23188} & {\scriptsize{}356.3 $\pm$ 64} & {\scriptsize{}61.067} & {\scriptsize{}344.592321}\tabularnewline
 
{\scriptsize{}orbital period {[}yrs{]}} & {\scriptsize{}81.2} & {\scriptsize{}\textendash{}} & {\scriptsize{}\textendash{}} & {\scriptsize{}\textendash{}} & {\scriptsize{}\textendash{}}\tabularnewline

{\scriptsize{}epoch of osculation {[}TT{]}} & {\scriptsize{}2015 Feb. 9.0} & {\scriptsize{}2014 Feb. 6.0} & {\scriptsize{}2014 Dec. 1.0} & {\scriptsize{}n.a.} & {\scriptsize{}2015 Feb. 9.2}\tabularnewline
\hline 
\end{tabular}
\end{table*}

Such a preliminary orbit (distant, Halley type comet) as well as the
low brightness of the object (20.1 \textendash{} 21.4 mag) probably
discouraged other observers and only three additional observations
were obtained on December 22, 2014, from iTelescope Observatory, Mayhill,
NM, USA, which previously observed this comet on November 28. This
poor quality orbit is copied on some Halley type comet lists\footnote{https://physics.ucf.edu/\textasciitilde{}yfernandez/cometlist.html}
and even mentioned in a recently published paper \citep{Yang_Ischiguro:2018}.

We may speculate that a small total number of observations and
large residuals obtained for the last three observations with respect
to the previously published orbit stimulated the staff at MPC (probably as a routine) to replace the previous orbit with the new one, based
on all 17 observations but with the eccentricity assumed to be 1.0.
This 'final' MPC orbit is presented in the second data column in Table
\ref{tab:various_preliminary_orbits}.

The parabolic orbit is dramatically different from the short period
elliptical one previously published. All elements are significantly
different but the most striking (at least for us) change is in the
perihelion passage date. While in the elliptical solution a comet
was observed nine month AFTER the perihelion and goes away, in the
parabolic solution it is observed 2.5 month BEFORE perihelion and
is approaching.

This poor quality orbit is reprinted in several Internet places, for
example at the Seiichi Yoshida's popular cometary catalogue\footnote{http://www.aerith.net/comet/catalog/2014W10/2014W10.html},
or on some Halley type comet lists\footnote{https://physics.ucf.edu/\textasciitilde{}yfernandez/cometlist.html}.
At the JPL Small Body database one might find also a parabolic orbit
for C/2014~W$_{10}$, based on all 17 observations with the mean
weighted residual of 0\farcs24, very similar to the 'final' MPC solution
(where mean residual is not given). The similarity is so good that
one might guess, that this is the same orbit as at MPC, but calculated
for a different osculation epoch. Surprisingly there is one strange
thing at JPL \textendash{} they provide uncertainties for all elements
INCLUDING ECCENTRICITY. Available is also a full 6$\times$6 covariance
matrix there so it seems that they OBTAINED not ASSUMED a parabolic
orbit.

It might be worth to mention, that from the set of first 14 observations
Japan amateur astronomer Kazuo Kinoshita obtained a completely different
but also parabolic orbit, perfectly fitting to  this subset of  observations\footnote{http://jcometobs.web.fc2.com/cmt/k14w10.htm}.
By putting $e=1.0$ he obtained $q=6.51927$ only, different angular
elements and the perihelion passage date even A YEAR LATER: $T=$2016
Aug. 17.267 TT with the mean residual of 0\farcs19.

All the above motivated us to check carefully what orbital information
can be obtained from these 17 observations available at MPC.

\section{New orbit determination}

First of all we observe, that in all previous solutions a perihelion
distance is large, what means that nongravitational forces might play
a marginal role in the case of this object so we decided to deal with
purely gravitational solutions only, which is also a typical approach for such a short observational arc.

Our first unresolved puzzle is how to obtain a Halley type orbit from
the first 14 observations. Instead, from these three day arc we obtained
a hyperbolic osculating orbit (solution A0, see the first column
of Table~\ref{tab:our_orbits}) which is highly incompatible (especially
from the point of view of the eccentricity) with that obtained at
MPC. The additional argument that the preliminary MPC orbit might
be significantly wrong is that additional three observations made
on December 22 drastically diverge by over two arc minutes. On the
contrary they are considerably closer to our solution A0, on a level
of 20\,arcsec.

Our next solution, named A2 and based on all 17~observations is presented in
the second column of Table~\ref{tab:our_orbits}. This orbit fits
very well to all observations, residua are presented in the first
column of Table~\ref{tab:our_orbits-residuals}. Since all these values are pretty small we did
not introduce any weighing. 

Can this solution be improved further? We decided to check whether
'catalogue debiasing' according to the procedure described by \citet{Farnocchia:2015}
change the orbital solution essentially.
According to our previous experiences these corrections, based on
averaged catalogue biases and approximating  unknown proper motions
are of little use in cometary astrometry, mainly due to intrinsic
errors of measuring diffuse object position, usually significantly
greater than these corrections. However in this case, where all residua
are so small (one may say: asteroidal) we decided to apply them, obtaining another solution,
named~D2.

We were really surprised obtaining an osculating eccentricity even
more hyperbolic, see the last column of Table\textasciitilde{}\ref{tab:our_orbits}.
Since the mean residual is slightly decreased, orbital element uncertainties
on the same level or smaller and all particular residua are pretty
small we decided to announce our solution D2 as the definitive orbit
of C/2014~W$_{10}$, provided no additional observations will be
found. Again, no observations weighing was applied. 

\begin{table*}
\caption{\label{tab:our_orbits}Osculating orbits of C/2014~W$_{10}$ PANSTARRS
calculated in this paper. Angular elements with respect to the ecliptic
of J2000.}

\begin{tabular}{cccc}
\hline 
Solution code & A0 & A2 & D2\tabularnewline
\hline 
number of observations & 14 & 17 & 17\tabularnewline

mean residual & 0\farcs19 & 0\farcs29 & 0\farcs24\tabularnewline
 
perihelion passage date {[}TT{]} & 2016 Jul. 15.6 $\pm$ 211 days & 2013 Jul. 3.8 $\pm$ 102 days & 2013 May 16.2 $\pm$ 42 days\tabularnewline
 
perihelion distance {[}AU{]} & 4.40 $\pm$ 2.26 & 7.575 $\pm$ 0.43 & 7.279 $\pm$ 0.40\tabularnewline
 
eccentricity &  1.400 $\pm$ 0.585 & 1.376 $\pm$ 0.379 & 1.653 $\pm$ 0.409\tabularnewline

inclination {[}deg{]} & 64.86 $\pm$ 18.8 & 89.14 $\pm$ 3.3 & 91.33 $\pm$ 3.0\tabularnewline

longitude of the ascending node {[}deg{]} & 38.74 $\pm$ 5.0 & 44.61 $\pm$ 0.8 & 45.11 $\pm$ 0.7\tabularnewline

argument of perihelion {[}deg{]} & 84.81 $\pm$ 19.4 & 339.41 $\pm$ 10.7 & 333.18 $\pm$ 7.4\tabularnewline

osculating heliocentric 1/a {[}AU$^{-1}${]} & \textendash 0.0907 $\pm$ 0.2349 & \textendash 0.0497 $\pm$ 0.0607 & \textendash 0.0897 $\pm$ 0.0744\tabularnewline

epoch of osculation {[}TT{]} & 2015 Feb. 9.0 & 2014 Feb. 6.0 & 2014 Dec. 1.0\tabularnewline
\hline 
\end{tabular}
\end{table*}

\begin{table*}
\caption{\label{tab:our_orbits-residuals}Osculating orbits of C/2014~W$_{10}$
calculated in this paper. Angular elements with respect to the ecliptic
of J2000.}

\begin{tabular}{cccccc}
\hline 
Date & Observatory  & \multicolumn{2}{c}{A2 solution residuals} & \multicolumn{2}{c}{D2 solution residuals}\tabularnewline
\hline 
(UT) & code & $\Delta\alpha$ & $\Delta\delta$ & $\Delta\alpha$ & $\Delta\delta$\tabularnewline

2014 11 25.25361  & F51 & \textendash{} 0\farcs38 & + 0\farcs05 & \textendash{} 0\farcs33 & \textendash{} 0\farcs01\tabularnewline
 
2014 11 25.26554  & F51 & \textendash{} 0\farcs39 & + 0\farcs32 & \textendash{} 0\farcs35 & + 0\farcs27\tabularnewline
 
2014 11 25.27744  & F51 & \textendash{} 0\farcs07 & \textendash{} 0\farcs11 & \textendash{} 0\farcs03 & \textendash{} 0\farcs16\tabularnewline
 
2014 11 25.28933  & F51 & + 0\farcs10 & \textendash{} 0\farcs03 & + 0\farcs13 & \textendash{} 0\farcs08\tabularnewline
 
2014 11 26.215967  & 568 & + 0\farcs42 & + 0\farcs12 & + 0\farcs34 & + 0\farcs18\tabularnewline

2014 11 26.217927  & 568 & + 0\farcs40 & + 0\farcs12 & + 0\farcs32 & + 0\farcs18\tabularnewline
 
2014 11 26.219888  & 568 & + 0\farcs34 & + 0\farcs10 & + 0\farcs27 & + 0\farcs16\tabularnewline

2014 11 28.182691  & H01 & + 0\farcs10 & + 0\farcs00 & \textendash{} 0\farcs03 & + 0\farcs13\tabularnewline
 
2014 11 28.187113  & H01 & + 0\farcs24 & + 0\farcs00 & + 0\farcs11 & + 0\farcs13\tabularnewline

2014 11 28.202180  & H01 & + 0\farcs22 & + 0\farcs06 & + 0\farcs08 & + 0\farcs19\tabularnewline

2014 11 28.210921  & H01 & + 0\farcs19 & \textendash{} 0\farcs23 & + 0\farcs06 & \textendash{} 0\farcs10\tabularnewline

2014 11 28.25174  & H06 & \textendash{} 0\farcs24 & \textendash{} 0\farcs14 & \textendash{} 0\farcs04 & \textendash{} 0\farcs30\tabularnewline

2014 11 28.25565  & H06 & \textendash{} 0\farcs21 & \textendash{} 0\farcs22 & \textendash{} 0\farcs02 & \textendash{} 0\farcs38\tabularnewline

2014 11 28.25955  & H06 & \textendash{} 0\farcs71 & \textendash{} 0\farcs11 & \textendash{} 0\farcs51 & \textendash{} 0\farcs27\tabularnewline
 
2014 12 22.08098  & H06 & + 0\farcs32 & + 0\farcs16 & + 0\farcs32 & + 0\farcs15\tabularnewline

2014 12 22.08407  & H06 & \textendash{} 0\farcs13 & + 0\farcs12 & \textendash{} 0\farcs13 & + 0\farcs12\tabularnewline

2014 12 22.08716  & H06 & \textendash{} 0\farcs19 & \textendash{} 0\farcs22 & \textendash{} 0\farcs19 & \textendash{} 0\farcs22\tabularnewline
\hline 
\end{tabular}
\end{table*}

\begin{table*}
\caption{\label{tab:our-original-orbits}Original barycentric orbits of C/2014~W$_{10}$.
Angular elements with respect to the ecliptic of J2000. }

\begin{tabular}{ccc}
\hline 
Corresponding osculating solution code & A2 & D2\tabularnewline
\hline 
perihelion passage date {[}TT{]} & 2013 Jul. 4.7  & 2013 May 15.7 \tabularnewline

perihelion distance {[}AU{]} & 7.579  & 7.276 \tabularnewline

eccentricity & 1.367  & 1.650 \tabularnewline

inclination {[}deg{]} & 89.11  & 91.35 \tabularnewline
 
longitude of the ascending node {[}deg{]} & 44.57  & 45.08 \tabularnewline
 
argument of perihelion {[}deg{]} & 339.53  & 333.15 \tabularnewline

epoch of osculation {[}TT{]} & 1863 Jan. 29.0 & 1894 Aug. 14.0\tabularnewline

original barycentric 1/a {[}AU$^{-1}${]} & \textendash0.0485  & \textendash0.0893 \tabularnewline
\hline 
\end{tabular}
\end{table*}

\section{Possible dynamical past of C/2014~W$_{10}$}

At first we would like to stress that due to large uncertainties of
the osculating (and therefore also the original) orbit parameters
nothing can be said about the dynamical past of this comet for sure.
However, according to very small obtained residua and large perihelion
distance (which makes significant nongravitational effects improbable)
we describe here the probable past dynamics of this object. \textbf{Our
main purpose is to show that similar cases should be treated in future
with greater care by more reliable preliminary orbit determination
and alerting observers about the importance of the object to initiate
more follow-up observations.}

\begin{figure}
\includegraphics[angle=270,width=0.95\columnwidth]{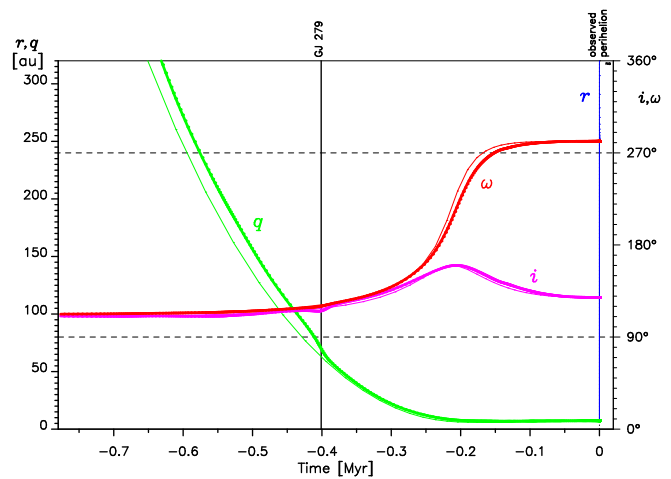}

\caption{\label{fig:Past-evolution}Past evolution of the nominal orbit of
C/2014~W$_{10}$ under the Galactic and stellar perturbations. Thin
lines depict the evolution with the stellar perturbations excluded.
Left vertical axis describes a heliocentric distance (blue line) and
an osculating perihelion distance (green line). Right vertical axis
describes angular elements in the galactic frame: an inclination (magenta
line) and am argument of perihelion (red line).}

\end{figure}

To obtain original orbit parameters of C/2014~W$_{10}$ along with
their uncertainties we performed a procedure similar to that performed
in our other papers, see for example \citet{kroli_dyb:2017}. In short,
we added 5000 virtual comets (VCs) to the nominal orbit obtained in
solutions A2 and D2 and followed numerically their motion backwards
to a heliocentric distance of 250~au. All planetary perturbations
as well as relativistic corrections were applied. Finally we changed
a reference frame to the barycentric one.

Elements of the nominal original orbits of C/2014~W$_{10}$ PANSTARRS obtained
from our solutions A2 and D2 are presented in Table~\ref{tab:our-original-orbits}. 
We do not list their large uncertainties (as useless) because it appeared that we are dealing with an extremely scattered VCs cloud. Instead, to illustrate how this swarm of VCs is dispersed we mention  that VCs reached a distance of 250\,au from the Sun in a wide range of nearly 400~years. This shows us how fundamentally this VCs-cloud differs from all the others we have studied so far. 
Therefore Table~\ref{tab:our-original-orbits} only gives orbital elements for nominal solutions (without uncertainties).
The original VCs swarm is also much more dispersed than the osculating cloud of VCs and it contains orbits with the original eccentricities in the range from $\sim$0.96 to more than 2.0.

\subsection{Nominal original orbit past evolution}

Just to illustrate the possible interstellar origin of C/2014~W$_{10}$ we studied the past evolution of the obtained nominal orbit of this comet. According to our nominal solution D2, comet C/2014~W$_{10}$ reached  a heliocentric distance of 250~au in 1894, moving towards the sun with a velocity of 9.29\,km~s$^{-1}$ from a direction $\alpha=14^{h}03^{m}$ and $\delta=-43\degr35'$. This direction is almost exactly at right angle to the solar apex
at $\alpha=18.5^{h}$ and $\delta=+30\degr$.

During the past 1.2 million years C/2014~W$_{10}$ travelled \textasciitilde{}10\,pc,
finally approaching our sun in 2014. Using the nominal original orbit
we have checked 3865 stars that can approach the sun closer than 10~pc
selected by \citet{Bailer-Jones:2018} from the Gaia DR2 catalogue
\citep{Gaia-DR2:2018} and additionally 3440 nearby stars (some overlap
of course) from the SIMBAD database \citep{SIMBAD-article:2000}.
None of these stars passed close enough to perturb C/2014~W$_{10}$
motion significantly or to direct it towards the sun from an Oort
cloud orbit. The orbital evolution of this comet during last 0.75\,Myrs is presented  in Fig.~\ref{fig:Past-evolution}. Thin lines show
the evolution without stellar perturbations while the thick ones show
the results of the full dynamical model. One can observe rather weak
perturbation from the star GJ~279 (HD~60532). This star passed approximately
0.75\,pc from this comet 0.4\,Myr ago at a relative velocity of 62\,km\,s$^{-1}$ with respect to the comet on its nominal orbit. 

Please note that this past evolution of C/2014~W$_{10}$ PANSTARRS is based on a extremely uncertain original orbit. On the other hand this orbit fits well (as well as all other VCs orbits) to our limited but precise observational data, so, cannot be ruled out.

\bibliographystyle{aa}
\bibliography{moja24}

\begin{thebibliography}{6}
\expandafter\ifx\csname natexlab\endcsname\relax\def\natexlab#1{#1}\fi

\bibitem[{{Bailer-Jones} {et~al.}(2018){Bailer-Jones}, {Rybizki}, {Andrae}, \&
  {Fouesneau}}]{Bailer-Jones:2018}
{Bailer-Jones}, C.~A.~L., {Rybizki}, J., {Andrae}, R., \& {Fouesneau}, M. 2018,
  A\&A, 616, A37

\bibitem[{{Farnocchia} {et~al.}(2015){Farnocchia}, {Chesley}, {Chamberlin}, \&
  {Tholen}}]{Farnocchia:2015}
{Farnocchia}, D., {Chesley}, S.~R., {Chamberlin}, A.~B., \& {Tholen}, D.~J.
  2015, Icarus, 245, 94

\bibitem[{{Gaia Collaboration} {et~al.}(2018){Gaia Collaboration}, {Brown},
  {Vallenari}, {Prusti}, {de Bruijne}, {Babusiaux}, {Bailer-Jones}, {Biermann},
  {Evans}, {Eyer}, \& et~al.}]{Gaia-DR2:2018}
{Gaia Collaboration}, {Brown}, A.~G.~A., {Vallenari}, A., {et~al.} 2018, A\&A,
  616, A1

\bibitem[{{Kr{\'o}likowska} \& {Dybczy{\'n}ski}(2017)}]{kroli_dyb:2017}
{Kr{\'o}likowska}, M. \& {Dybczy{\'n}ski}, P.~A. 2017, MNRAS, 472, 4634

\bibitem[{{Wenger} {et~al.}(2000){Wenger}, {Ochsenbein}, {Egret}, {Dubois},
  {Bonnarel}, {Borde}, {Genova}, {Jasniewicz}, {Lalo{\"e}}, {Lesteven}, \&
  {Monier}}]{SIMBAD-article:2000}
{Wenger}, M., {Ochsenbein}, F., {Egret}, D., {et~al.} 2000, A\&A Supplement
  Series, 143, 9

\bibitem[{{Yang} \& {Ishiguro}(2018)}]{Yang_Ischiguro:2018}
{Yang}, H. \& {Ishiguro}, M. 2018, \apj, 854, 173

\end{thebibliography}

\end{document}